\documentclass[aps,prd,twocolumn,groupedaddress,showpacs]{revtex4}

\usepackage{graphicx}
\usepackage{dcolumn}
\usepackage{bm}
\usepackage{amsmath}
\usepackage{epstopdf}
\usepackage{amsfonts}
\usepackage{amssymb}
\usepackage{hyperref}
\usepackage{xcolor}
\usepackage{float}

\usepackage{natbib}

\providecommand{\U}[1]{\protect\rule{.1in}{.1in}}

\newcommand{\baa}{\begin{align}}
\newcommand{\eaa}{\end{align}}

\newcommand{\be}{\begin{equation}}
\newcommand{\ee}{\end{equation}}
\newcommand{\bea}{\begin{eqnarray}}
\newcommand{\eea}{\end{eqnarray}}

\begin{document}

\title{Growth index and statefinder diagnostic of Oscillating Dark Energy}

\author{Grigoris Panotopoulos}
\affiliation{Centro de Astrof{\'i}sica e Gravita{\c c}{\~a}o, Instituto Superior T{\'e}cnico-IST,
Universidade de Lisboa-UL, Av. Rovisco Pais, 1049-001 Lisboa, Portugal}
\email{grigorios.panotopoulos@tecnico.ulisboa.pt}

\author{\'Angel Rinc\'on}
\affiliation{Instituto de F\'{i}sica, Pontificia Universidad Cat\'{o}lica de Chile, \mbox{Avenida Vicu\~na Mackenna 4860, Santiago, Chile.}}
\email{arrincon@uc.cl}

\date{\today}

\begin{abstract}
We study is some detail the Cosmology of Oscillating Dark Energy described by concrete equations-of-state investigated recently in the literature. In particular, at the background level we compute the statefinder parameters, while at the level of linear cosmological perturbations we compute the growth index $\gamma$ as well as the combination parameter $A=f \sigma_8$. The comparison with $\Lambda$CDM is made as well.
\end{abstract}

\maketitle

\section{Introduction}

In the end of the 1990s the breakthrough in science was the discovery that the Universe was expanding in an accelerating rate \cite{SN1,SN2}. 
Current cosmological and astrophysical observational data indicate that the Universe is spatially flat dominated by dark energy \cite{turner}
the origin and nature of which still remain a mystery. The concordance cosmological model, which is based on cold dark matter and a cosmological constant (or $\Lambda$CDM model), is considered to be the simplest choice and the most economical model. It is characterized by a single parameter, and it is in excellent agreement with current data. Despite its success, however, it suffers from the cosmological constant problem \cite{weinberg}. That is why other possibilities have been explored in the literature over the years, which in general fall into broad classes, namely either dynamical or geometrical models of dark energy. In the first class a new dynamical field is introduced to accelerate the Universe \cite{Copeland:2006wr}, while in the second case an alternative theory of gravity is assumed to modify Einstein's General Relativity at cosmological scales \cite{dgp,HS,starobinsky}.

Although a fundamental description based on the Lagrangian formalism is certainly the ideal one, a phenomenological description is simpler and more convenient. In such a description dark energy is viewed as a perfect fluid with a time varying equation-of-state $w(a)$, with $a$ being the scale factor. In the past in \cite{Nesseris:2004wj} the authors compared several dark energy parameterizations against supernovae data, and concluded that models that cross the $w=-1$ barrier have a better fit to data. In another more recent work it was shown that the cosmic acceleration may have slowed down recently \cite{Magana:2014voa}. Oscillating Dark Energy (ODE) is a class of dark energy parameterization that has been studied by several authors and in different contexts \cite{ODE1,extra,ODE2,ODE3,ODE4,ODE5}, and more recently in \cite{pace,saridakis}, during the last 15 years or so. A couple of reasons why studying this particular class of models is interesting are the following: In \cite{ODE1} it was demonstrated that ODE
may alleviate the coincidence problem, while in \cite{ODE2} it was shown that unifying the current cosmic acceleration with the inflationary Universe was possible. One of the first papers along these lines was the work of \cite{extra}, in which the graceful exit of cosmological inflation and reheating of the Universe were studied in the framework of dynamical relaxation of a bare cosmological constant.

It is well-known that many dark energy models predict very similar expansion histories, and therefore all of them are still in agreement with current data. It thus becomes clear that in order to discriminate between different dark energy models it is necessary to introduce and study new quantities appropriately defined. To this end one option would be to study the so-called statefinder parameters, $r,s$, defined as follows \cite{Sahni:2002fz,Alam:2003sc}
\begin{eqnarray}
r & = & \frac{\dddot{a}}{a H^3} \\
s & = & \frac{r-1}{3 (q-\frac{1}{2})}
\end{eqnarray}
where the dot denotes differentiation with respect to the cosmic time $t$, $H=\dot{a}/a$ is the Hubble parameter, and $q=- \ddot{a}/(a H^2)$ is the decelerating parameter. We see that the statefinder parameters are expressed in terms of the third derivative of the scale factor with respect to the cosmic time, contrary to the Hubble parameter and the decelerating parameter, that are expressed in terms of the first and the second time derivative of the scale factor respectively. It is easy to check that for the $\Lambda$CDM model the statefinder parameters take constant values, $r=1,s=0$. These parameters may be computed within a certain model, their values can be extracted from future observations \cite{SNAP1,SNAP2}, and the statefinder diagnostic has been applied to several dark energy models \cite{diagnostics1,diagnostics2,diagnostics3,diagnostics4,diagnostics5}.
As we will see later on, $r,s$ can be very different from one model to another even if they predict very similar expansion histories. 

Alternatively, one may investigate how matter perturbations evolve with time. Its evolution depends both on the expansion history and on the speed of sound of the dark energy model at hand, which differs from one model to another. In particular, a quantity that has been studied a lot over the years is the so called growth index $\gamma$, introduced in \cite{Wang:1998gt} and to be defined later on, and for the $\Lambda$CDM model has been found to be $\gamma_{\Lambda\text{CDM}} = 6/11 \simeq 0.55$ \cite{Nesseris:2007pa,domenico,latino}. 

It is the goal of the present article to further study the Cosmology of the Oscillating Dark energy models studied recently \cite{pace,saridakis} in more detail along the lines mentioned before. Our work is organized as follows: after this introduction, we present the theoretical framework in the second section, and we present our numerical results in section 3. Finally we conclude in the last section.

\section{Theoretical framework}

Here we introduce all the necessary ingredients, first for the background and then for the evolution of linear cosmological perturbations.

\subsection{Background evolution}

The time dependence of the scale factor is determined by the first and the second Friedmann equations of a flat Robertson-Walker metric
\begin{eqnarray}
H^2 & = & \frac{8 \pi G \rho}{3} \\
\dot{H} & = & - 4 \pi G (\rho + p)
\end{eqnarray}
where $G$ is Newton's constant, $\rho = \rho_m + \rho_r + \rho_X$ is the total energy density, $p = p_m + p_r + p_X$ is the total pressure, and the index $m,r,X$ denotes the fluid component of matter, radiation and dark energy respectively. What is more, assuming no interaction between the fluid components, the continuity equation for each fluid is given by
\begin{equation}
\dot{\rho_A} + 3 H \rho_A (1+w_A) = 0
\end{equation}
where the index $A$ takes 3 values, $A=m,r,X$. The equation-of-state parameter for radiation is $w_r=1/3$, for matter $w_m=0$, while for dark energy we shall assume some parameterization where its equation-of-state will be a certain function of the red-shift $1+z=a_0/a$, with $a_0$ being the present value of the scale factor $a$. Finally, for later use we define the normalized energy density for each fluid, $\Omega_A \equiv \rho_A/\rho$. Thus, the first Friedmann equation now plays the role of the constraint $\sum_A \Omega_A \equiv 1$. 

The cosmological equations together with the definitions allow us to compute both the deceleration parameter $q$ and the first statefinder parameter $r$ as functions of the red-shift, which are found to be
\begin{eqnarray}
q(z) & = & -1+(1+z) \frac{E'(z)}{E(z)} \\
r(z) & = & q(z) (2q(z)+1) + (1+z) q'(z)
\end{eqnarray}
where $E(z)=H(z)/H_0$ is the dimensionless Hubble parameter versus red-shift, with $H_0=100 \: h (\text{km  sec}^{-1}) / \text{Mpc}$ being the Hubble constant, and the prime denotes differentiation with respect to red-shift. The second statefinder parameter $s(z)$ can be computed once $q(z),r(z)$ are known. It is easy to verify that the expressions above for $\Lambda$CDM give $r=1,s=0$. For matter and radiation the normalized energy densities are given by
\begin{eqnarray}
\Omega_m(z) & = & \frac{\Omega_{m,0} (1+z)^3}{E(z)^2} \\
\Omega_r(z) & = & \frac{\Omega_{r,0} (1+z)^4}{E(z)^2}
\end{eqnarray}
where $\Omega_{m,0}, \Omega_{r,0}$ are their present values.
Finally, for a given dark energy parameterization $w(z)$, the dimensionless Hubble parameter $E(z)=H(z)/H_0$ is given by
\begin{widetext}
\begin{equation}
E(z) = \sqrt{\Omega_{m,0} (1+z)^3 + \Omega_{r,0} (1+z)^4 + (1-\Omega_{m,0}-\Omega_{r,0}) F(z)}
\end{equation}
\end{widetext}
where $\Omega_{r,0}=9 \times 10^{-5}$ \cite{Magana:2014voa}, and where the function $F(z)$ is computed once the dark energy equation-of-state is given \cite{Nesseris:2004wj}
\begin{equation}
F(z) = \exp\left(3 \int_0^z \: dx \: \frac{1+w(x)}{1+x} \right)
\end{equation}
We see that $F(0)=1=E(0)$, as they should, since the constraint $\sum_A \Omega_A \equiv 1$ should be always satisfied, as we have already mentioned.

\subsection{Cosmological perturbations}

In this subsection we briefly present linear cosmological perturbation theory \cite{Mukhanov:2005sc,dePutter:2010vy}, following closely \cite{Albarran:2016mdu,panotopoulos}. Only scalar perturbations are relevant to structure formation, and assuming vanishing anisotropic stress tensor for the fluid components, the metric is characterized by a single Bardeen potential $\Psi(\eta, \vec{x})$
\begin{equation}
ds^2 = a(\eta)^2 [-(1+2 \Psi) d\eta^2 + (1-2 \Psi) \delta_{i j} dx^i dx^j]
\end{equation}
with $d\eta=dt/a$ being the conformal time. On the one hand, the perturbed Einstein's equations $\delta G_{\mu \nu}=8 \pi G \delta T_{\mu \nu}$ give rise to the differential equations for $\Psi$, which in Fourier space take the form \cite{Albarran:2016mdu}
\begin{align}
3 \mathcal{H} (\mathcal{H} \Psi+\Psi')+k^2 \Psi & =  -4 \pi G a^2 \delta \rho \\
\mathcal{H} \Psi+\Psi' & =  -4 \pi G a^2 (\rho+p) v \\
\Psi'' + 3 \mathcal{H} \Psi'+\Psi (\mathcal{H}^2+2 \mathcal{H}') & =  4 \pi G a^2 \delta p
\end{align}
where $\delta p=\sum_A \delta p_A$ is the total pressure perturbation, $\delta \rho=\sum_A \delta \rho_A$ is the total energy density perturbation, and
$(1+w) v = \sum_A (1+w_A) \Omega_A v_A$ is the total peculiar velocity potential \cite{Albarran:2016mdu}. On the other hand, the stress-energy tensor conservation for each fluid provides us with additional differential equations for the peculiar velocity potential $v$ as well as the density contrast $\delta_A = \delta \rho_A/\rho_A$. Defining the total density contrast $\delta = \delta \rho/\rho=\sum_A \Omega_A \delta_A$ one finally obtains the following equations for the metric perturbation \cite{Albarran:2016mdu}
\begin{eqnarray}
\Psi_x + \Psi \left(1+\frac{k^2}{3 \mathcal{H}^2} \right) & = & -\frac{\delta}{2} \\
\Psi_x + \Psi & = &  -\frac{3}{2} \mathcal{H} v (1+w)
\end{eqnarray}
where now $x=-\ln(1+z)$ is introduced as the independent variable, so that for any perturbation $A' = \mathcal{H} A_x$.
Additionally, one obtains for the fluid perturbations $\delta$ and $v$ the following equations \cite{Albarran:2016mdu}
\begin{align}
(\delta_r)_x  & =  \frac{4}{3} \left(3 \Psi_x + v_r \frac{k^2}{\mathcal{H}} \right)   
\\
(v_r)_x & =  -\frac{1}{\mathcal{H}} \left(\Psi + \frac{\delta_r}{4} \right)  
\\
(\delta_m)_x  & =  3 \Psi_x + v_m \frac{k^2}{\mathcal{H}} 
\\
(v_m)_x & =  - \left(v_m + \frac{\Psi}{\mathcal{H}} \right) 
\\
(\delta_X)_x & =  3 (w_X-c_{X,s}^2) \delta_X + (1+w_X) 
\nonumber
\\
& \times \Bigl[3 \Psi_x + v_X \left(\frac{k^2}{\mathcal{H}} + 9 \mathcal{H} (c_{X,s}^2-c_{X,a}^2) \right) \Bigl] 
\\
(v_X)_x & =  v_X (3 c_{X,s}^2-1)-\frac{1}{\mathcal{H}} \left(\Psi +\delta_X \frac{c_{X,s}^2}{1+w_X}  \right)
\end{align}
for all three fluid components. The adiabatic speed of sound $c_{A,a}^2$ is defined by \cite{dePutter:2010vy,Albarran:2016mdu}
\begin{equation}
c_{A,a}^2 = \frac{\dot{p_A}}{\dot{\rho_A}}=w_A-\frac{\dot{w_A}}{3 H (1+w_A)}
\end{equation}
while the effective speed of sound in the rest frame of the fluid $c_{A,s}^2$ is defined by \cite{dePutter:2010vy,Albarran:2016mdu}
\begin{equation}
\delta p_A = c_{A,s}^2 \delta \rho_A - 3 \mathcal{H} (1+w_A) \rho_A v_A (c_{A,s}^2-c_{A,a}^2)
\end{equation}
with $\mathcal{H} a = da/d\eta$ being the conformal Hubble parameter, and the prime denotes differentiation with respect to the conformal time. It is trivial to compute the sound speeds for matter and radiation, $c_{m,a}^2=0=c_{m,s}^2$ and $c_{r,a}^2=1/3=c_{r,s}^2$, respectively, while for dark energy, following \cite{saridakis}, we have taken $c_{X,s}^2=1$.

Finally, the system of coupled differential equations must be supplemented with the appropriate initial conditions. Single-field inflationary models predict adiabatic initial conditions
\begin{equation}\label{adiabatic}
\frac{\delta_i}{1+w_i} = \frac{\delta_j}{1+w_j}
\end{equation}
for any two fluids $i,j$. Therefore one obtains the following initial conditions for the peculiar velocity potentials \cite{Albarran:2016mdu}
\begin{equation}
v_{A,ini} = \frac{\delta_{ini}}{4 \mathcal{H}_{ini}}
\end{equation}
while for the density contrasts one obtains the initial conditions
\begin{equation}
\delta_{A,ini} = \frac{3}{4} (1+w_{A,ini}) \: \delta_{ini}
\end{equation}
Finally, for the relevant cosmological parameters, such as $H_0, \Omega_{m,0}$ etc, we have used the results of \cite{saridakis} shown in Tables 2, 3 and 4 of that work. 

The growth index $\gamma$ is defined through the relation below \cite{Linder1,Linder2,Linder3}, and for more recent discussions see e.g. \cite{Ballesteros:2008qk,growth1,growth2,growth3}
\begin{equation}
\frac{d (\ln\delta_m)}{d (\ln a)} = f = \Omega_m^\gamma
\end{equation}
Therefore, we first integrate the full system of coupled perturbations starting to follow their time evolution from the radiation dominated era where $z_{ini}=10^6$ or $x_{ini}=-13.81$, then we compute the function $f$ from the matter energy density contrast, and finally the growth index can be computed by
\begin{equation}
\gamma = \frac{\ln(f)}{\ln(\Omega_m)}
\end{equation}

There are a few publicly available computer codes that could be used to integrate the equations for the perturbations \cite{cmbfast,camb,cmbeasy,doran}. In this work, however, since we are not interested in the temperature anisotropies, we prefer to integrate the equations using a Wolfram Mathematica \cite{wolfram} file, as it was done in \cite{leandros}.

\section{Numerical results}

In this section we analyse the following 3 oscillatory equation-of-state parameters that were recently investigated in \cite{saridakis}
\begin{eqnarray}
w_I(z) & = & w_0 + b \bigl[1-\cos[\ln(1+z)] \bigl] \\
w_{II}(z) & = & w_0 + b \sin[\ln(1+z)] \\
w_{III}(z) & = & w_0 + b \Bigl[\frac{\sin(1+z)}{1+z}-\sin 1\Bigl]
\end{eqnarray}
where $w_0=w(0)$ is their today's value. For these parameterizations, the function $F(z)$ that determines the dimensionless
Hubble parameter is found to be
\begin{align}
F(z) &= (1+z)^{3(1+b+w_0)} e^{-3b \sin(\ln(1+z))}
\end{align}
for the first model,
\begin{align}
F(z) &= (1+z)^{3(1+w_0)} e^{3b(1-\cos \ln(1+z))}
\end{align}
for the second model, and
\begin{align}
F(z) = F_0^b e^{3b \bigl[\text{Ci}(1+z) -\frac{\sin(1+z)}{(1+z)} \bigl]} (1+z)^{3(1+w_0-b \sin(1))}
\end{align}
for the third model, where the cosintegral function is defined as follows: 
\begin{align}
\text{Ci}(z) \equiv - \int^{\infty}_{z} \mathrm{d}t \frac{\cos(t)}{t}
\end{align}
and we have introduced the real parameter $F_0 = e^{3(\sin(1) - \text{Ci}(1))}$.

Using several observational data, such as supernovae type Ia, BAO distance measurements, weak gravitational lensing, CMB observations etc, the authors of \cite{saridakis} found the values shown in Table I.

\begin{figure*}[t!]
\centering
\includegraphics[width=0.49\textwidth]{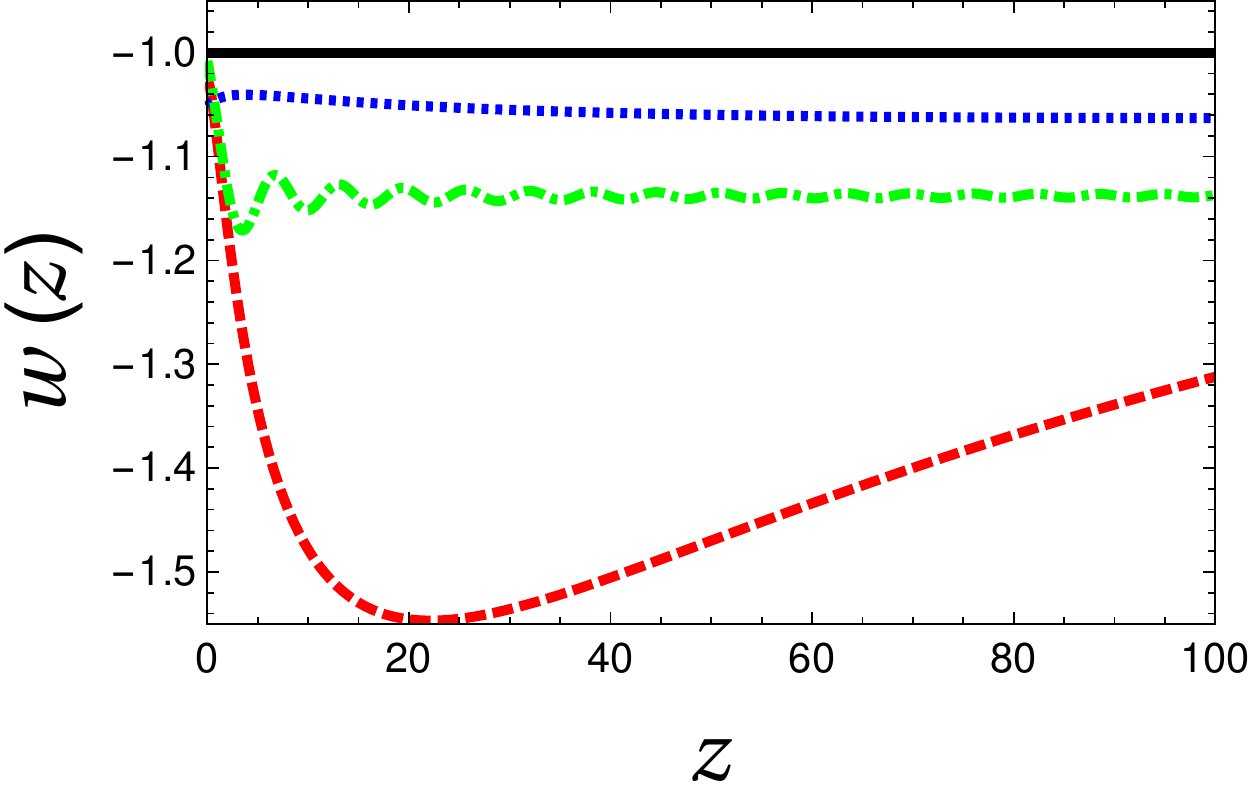} \
\includegraphics[width=0.49\textwidth]{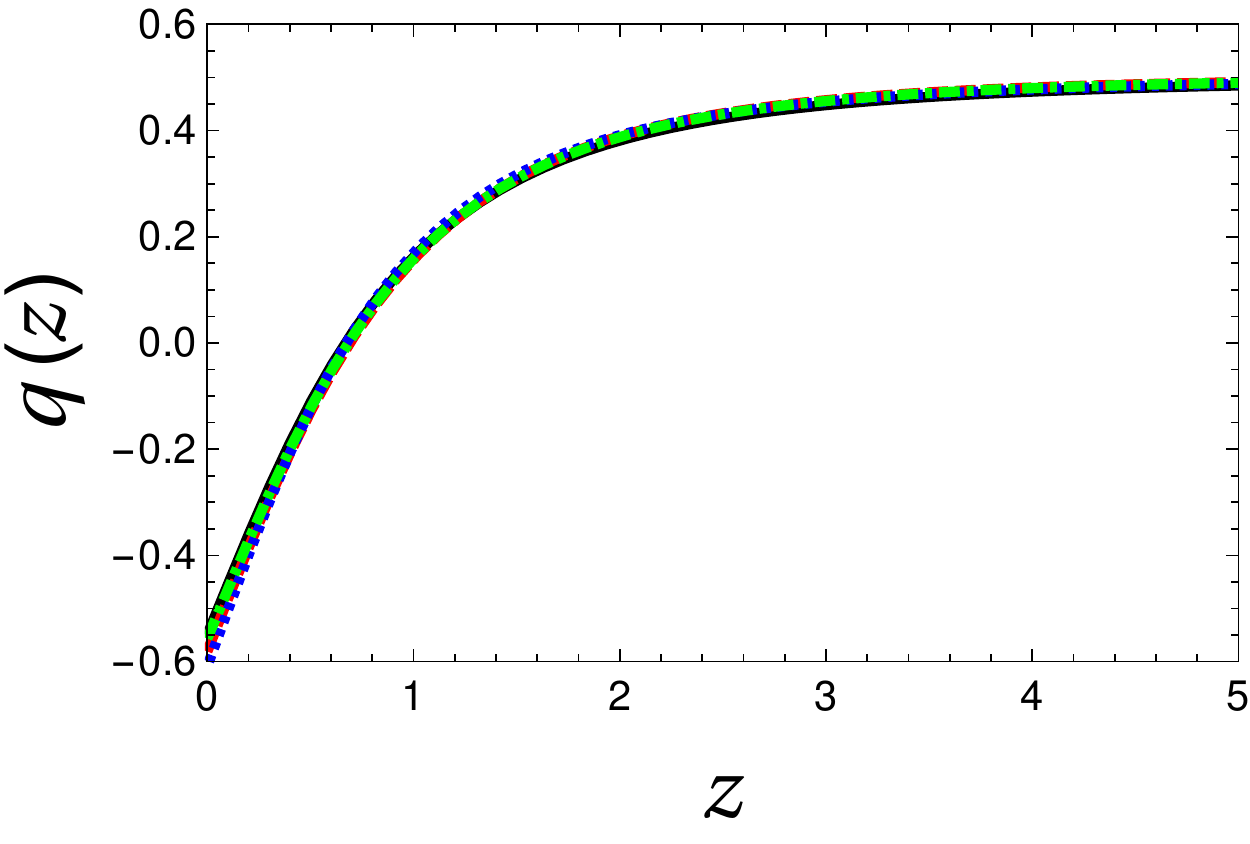}
\caption{
{\bf Left panel:} 
Equation of state parameter versus red-shift for the $\Lambda$CDM model (solid black line), for the first model (dashed red line), for the second model (dotted blue line) and for the third model (dotted dashed green line).
{\bf Right panel:} 
Deceleration parameter $q$ versus red-shift for the $\Lambda$CDM model (solid black line), for the first model (dashed red line), for the second model (dotted blue line) and for the third model (dotted dashed green line).
}
\label{fig:1}
\end{figure*}

\begin{figure*}[t!]
\centering
\includegraphics[width=0.49\textwidth]{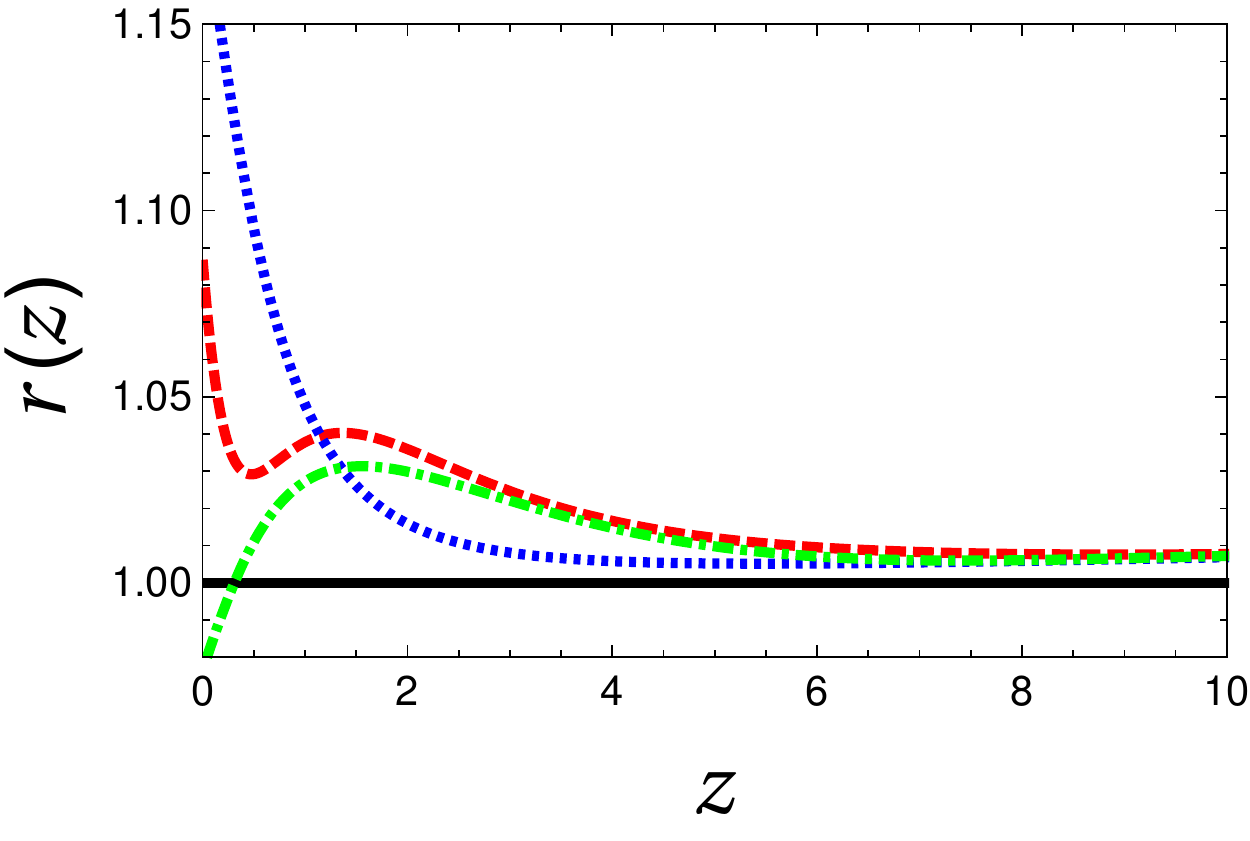} \
\includegraphics[width=0.49\textwidth]{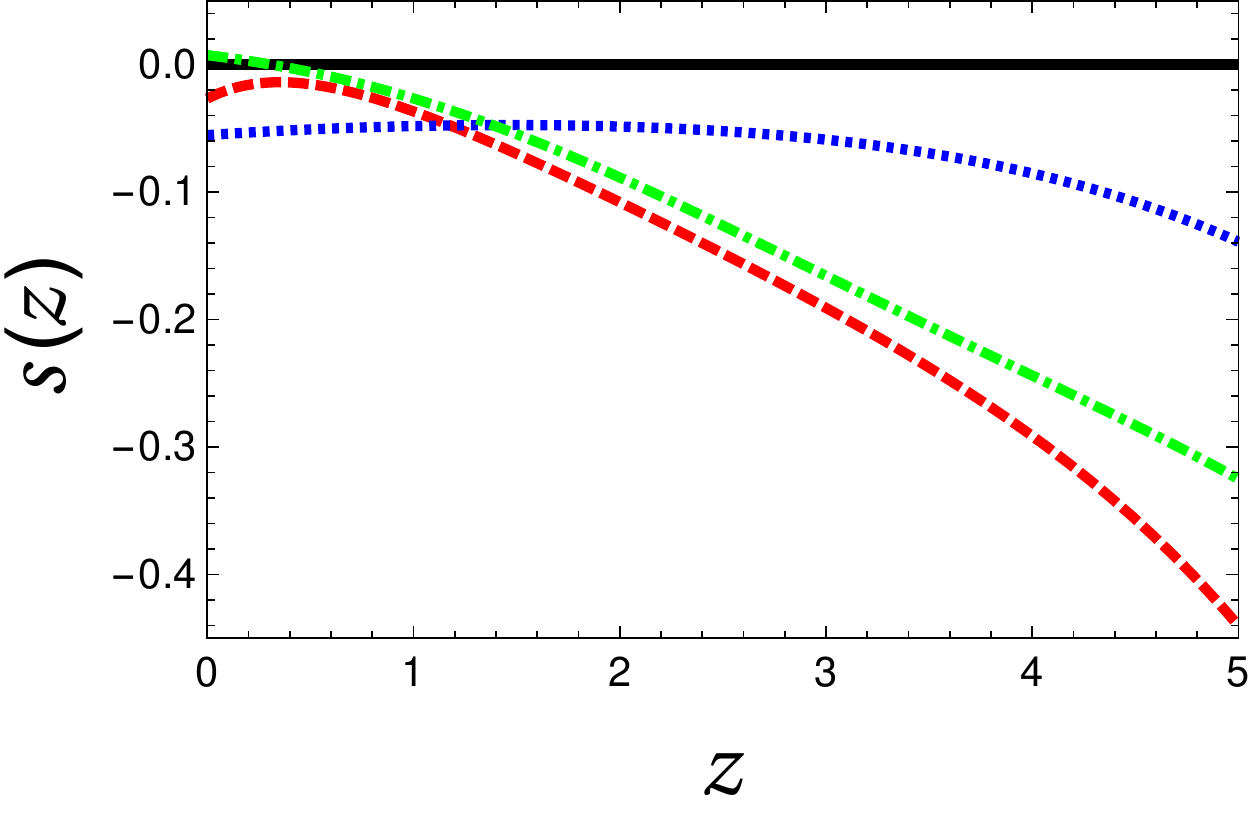}
\caption{
{\bf Left panel:} The first statefinder parameter $r$ versus red-shift for the $\Lambda$CDM model (solid black line), for the first model (dashed red line), for the second model (dotted blue line) and for the third model (dotted dashed green line). {\bf Right panel:} The second statefinder parameter $s$ versus red-shift for the $\Lambda$CDM model (solid black line), for the first model (dashed red line), for the second model (dotted blue line) and for the third model (dotted dashed green line).
}
\label{fig:2}
\end{figure*}

The equation of state and the deceleration parameter versus red-shift  are shown in Fig. \ref{fig:1} for all 3 models. We see that a) the equation-of-state in all 3 cases always remains in the range below the $-1$ barrier, and b) all models predict the same decelerating parameter as a function of red-shift. The first and second statefinder parameters $r$ and $s$ versus red-shift are shown in Fig. \ref{fig:2} for all 3 models. We see that the second statefinder parameter at low red-shift approaches that of $\Lambda$CDM, while at large $\Lambda$CDM deviates significantly from the standard behaviour. The opposite holds for the first parameter.

\begin{figure*}[ht]
\centering
\includegraphics[width=0.32\textwidth]{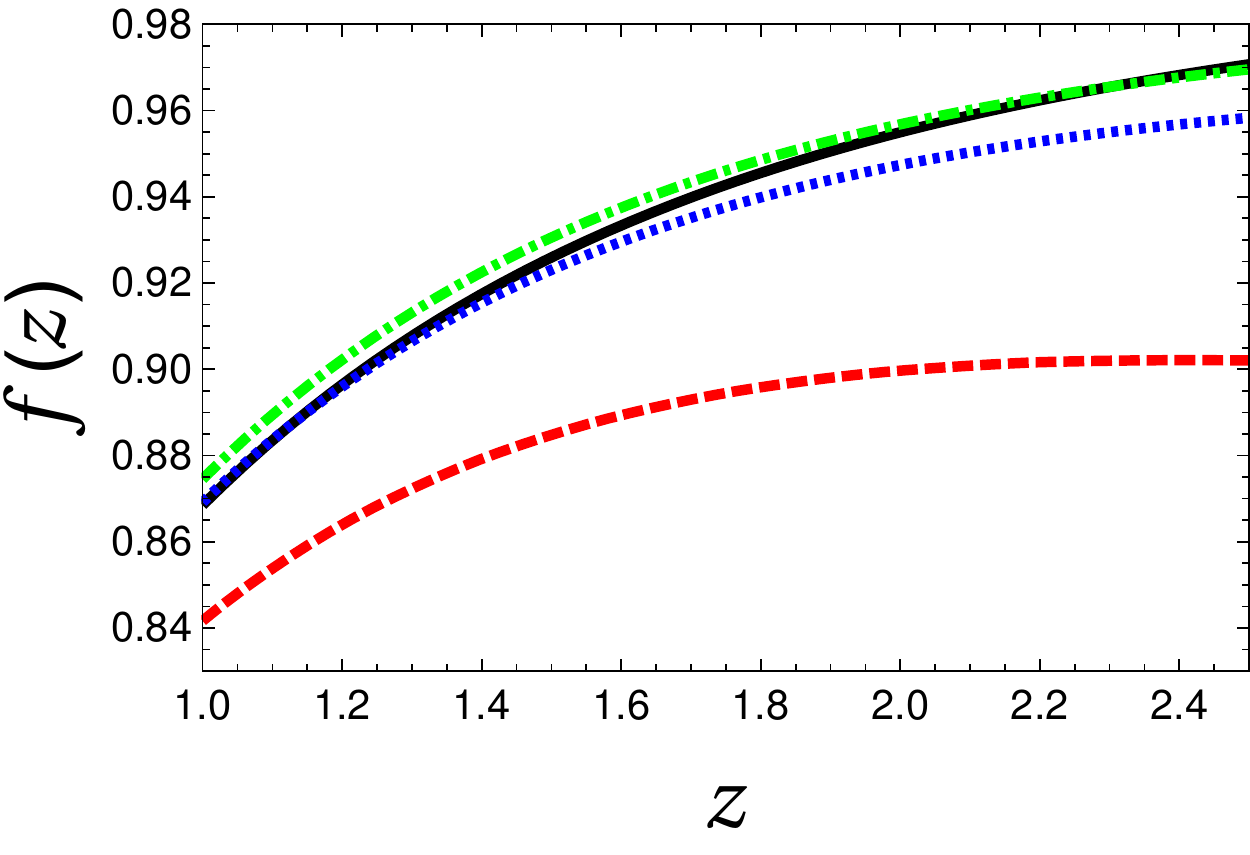}   \
\includegraphics[width=0.32\textwidth]{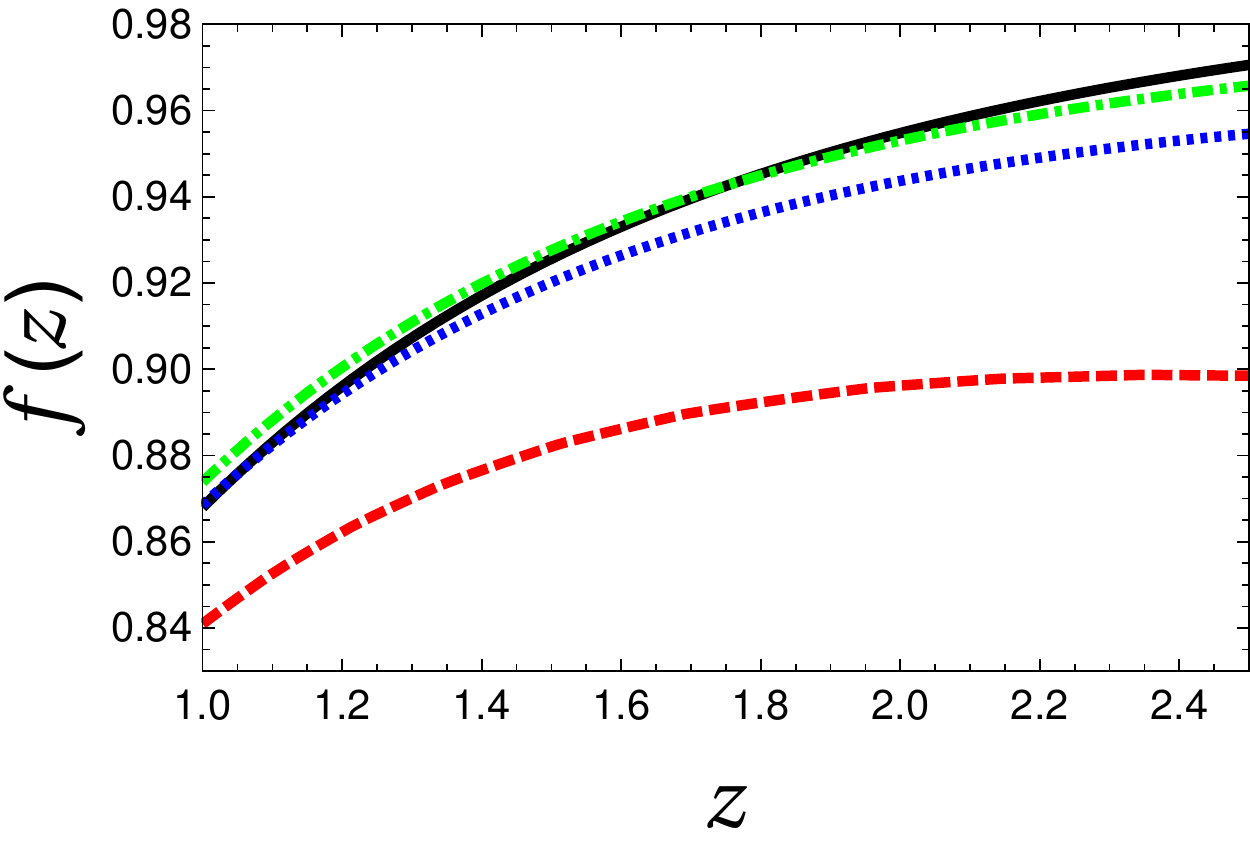}  \
\includegraphics[width=0.32\textwidth]{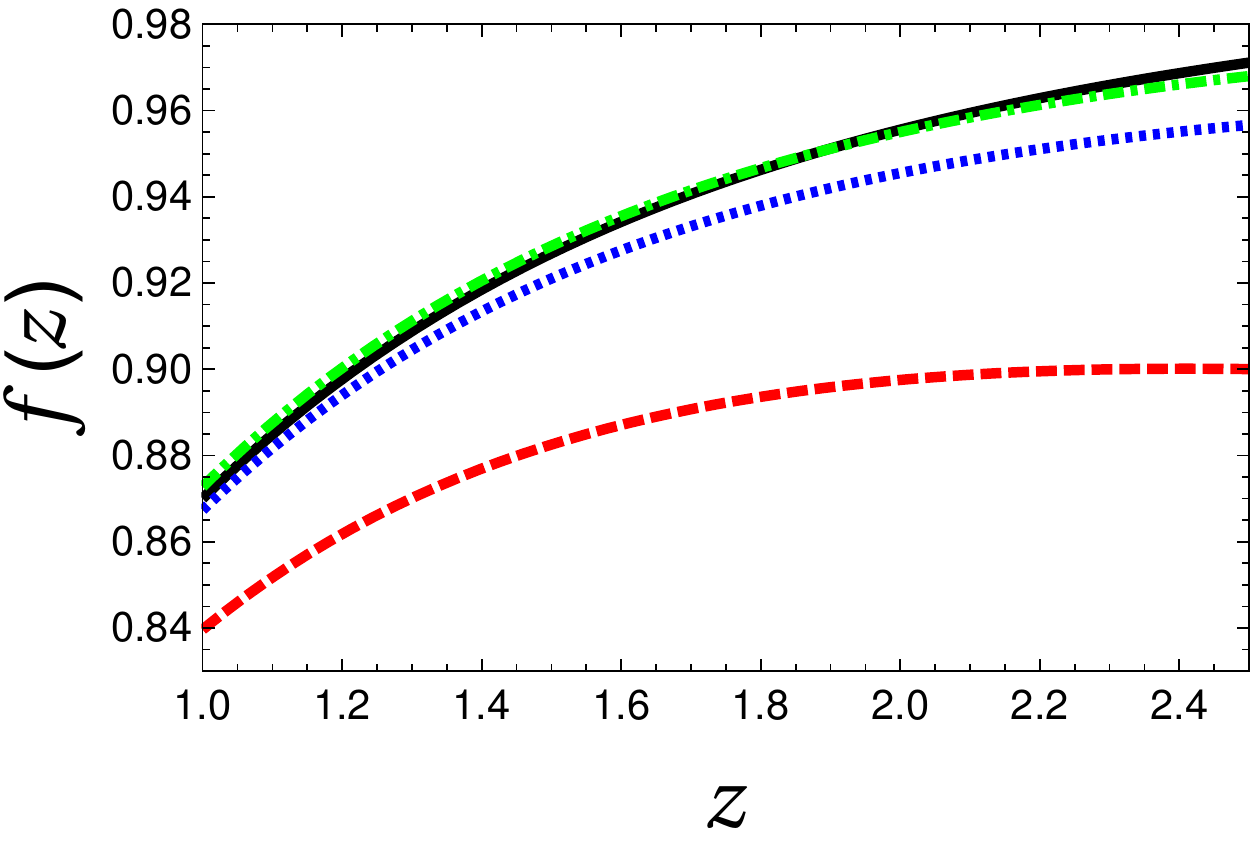} \

\medskip

\includegraphics[width=0.32\textwidth]{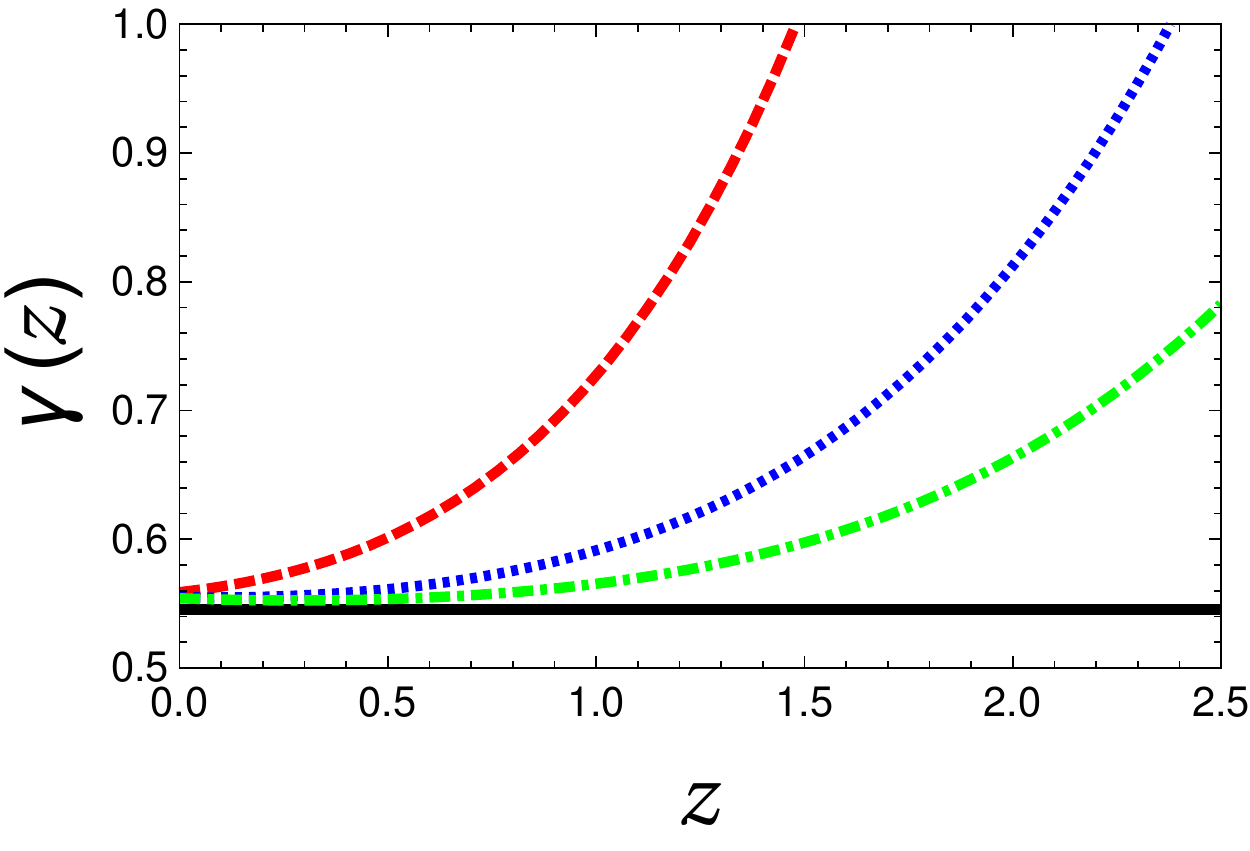} \
\includegraphics[width=0.32\textwidth]{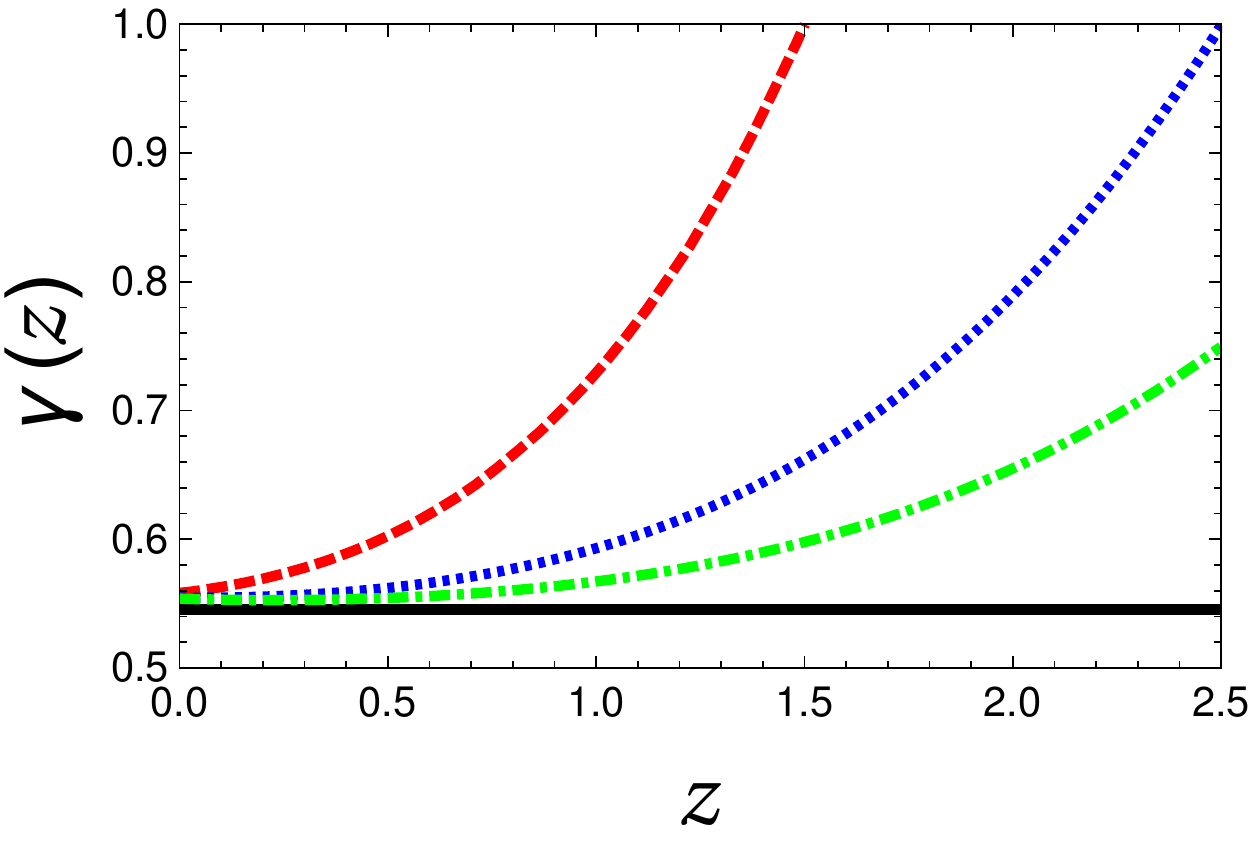} \
\includegraphics[width=0.32\textwidth]{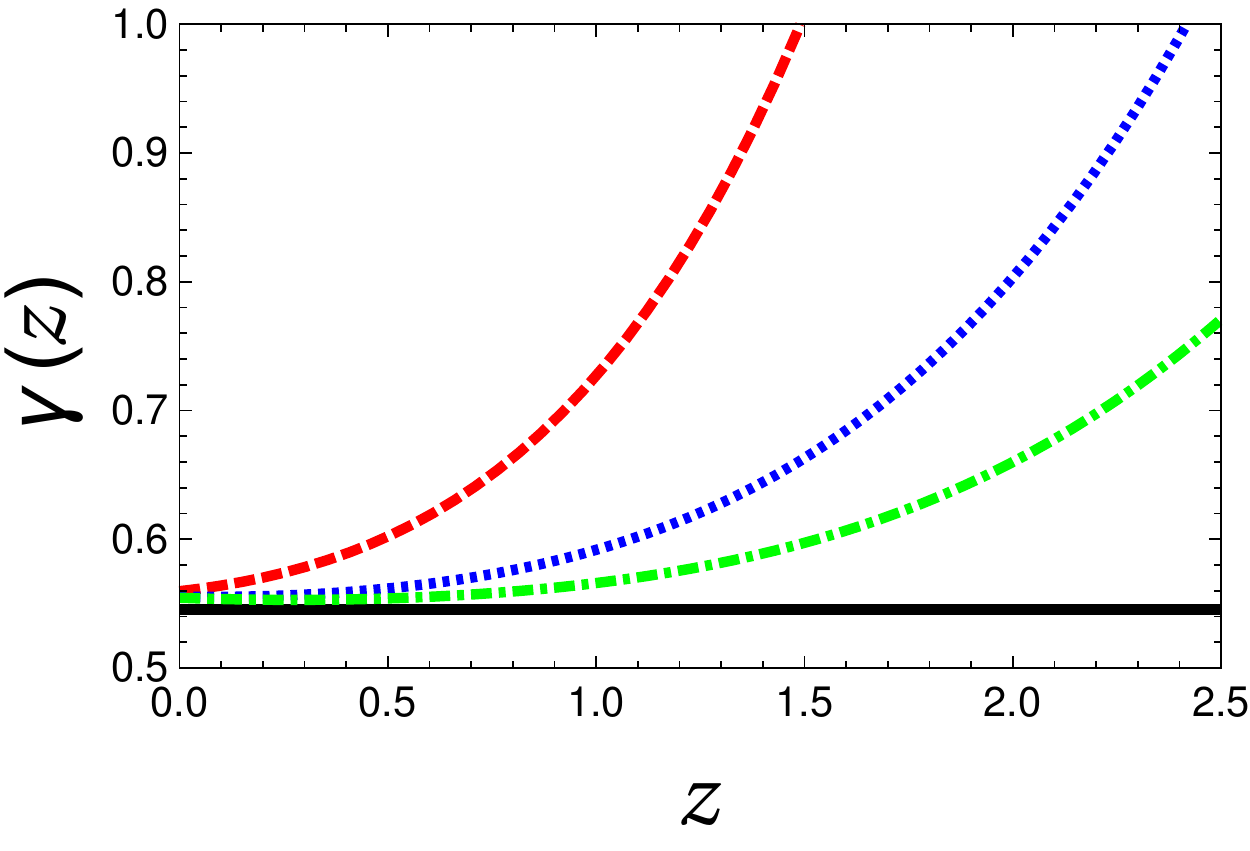}

\caption{
The evolution of functions $f$ and $\gamma$ versus red-shift $z$ for the three models.
The panels in the first (left), second (center) and third (right) column describe the functions $\{f(z), \gamma(z)\}$ for the model I, II and III respectively. 
We show the $\Lambda$CDM model (solid black line) and three different cases: i) $k = 2 \times 10^{-3}$ h Mpc$^{-1}$ (dashed red line), ii) $k = 4 \times 10^{-3}$ h Mpc$^{-1}$  (dotted blue line) and iii) for $k = 6 \times 10^{-3}$ h Mpc$^{-1}$  (dotted dashed green line).
}
\label{fig:3}
\end{figure*}

Next we consider linear cosmological perturbations and the evolution of the functions $\{f(z), \gamma(z)\}$. We show the solution for the three cases studied in Fig. \ref{fig:3} for three different scales $k$ of the linear regime, of the order of $10^{-3}$ Mpc. For comparison we show in the same plot (solid black curves) the corresponding quantities of the $\Lambda$CDM model. We see that i) in all three models both $f$ and $\gamma$ increase with red-shift, and ii) as the wave number $k$ increases the curves approach the curve corresponding to $\Lambda$CDM.

\begin{figure}[ht!]
\centering
\includegraphics[width=\linewidth]{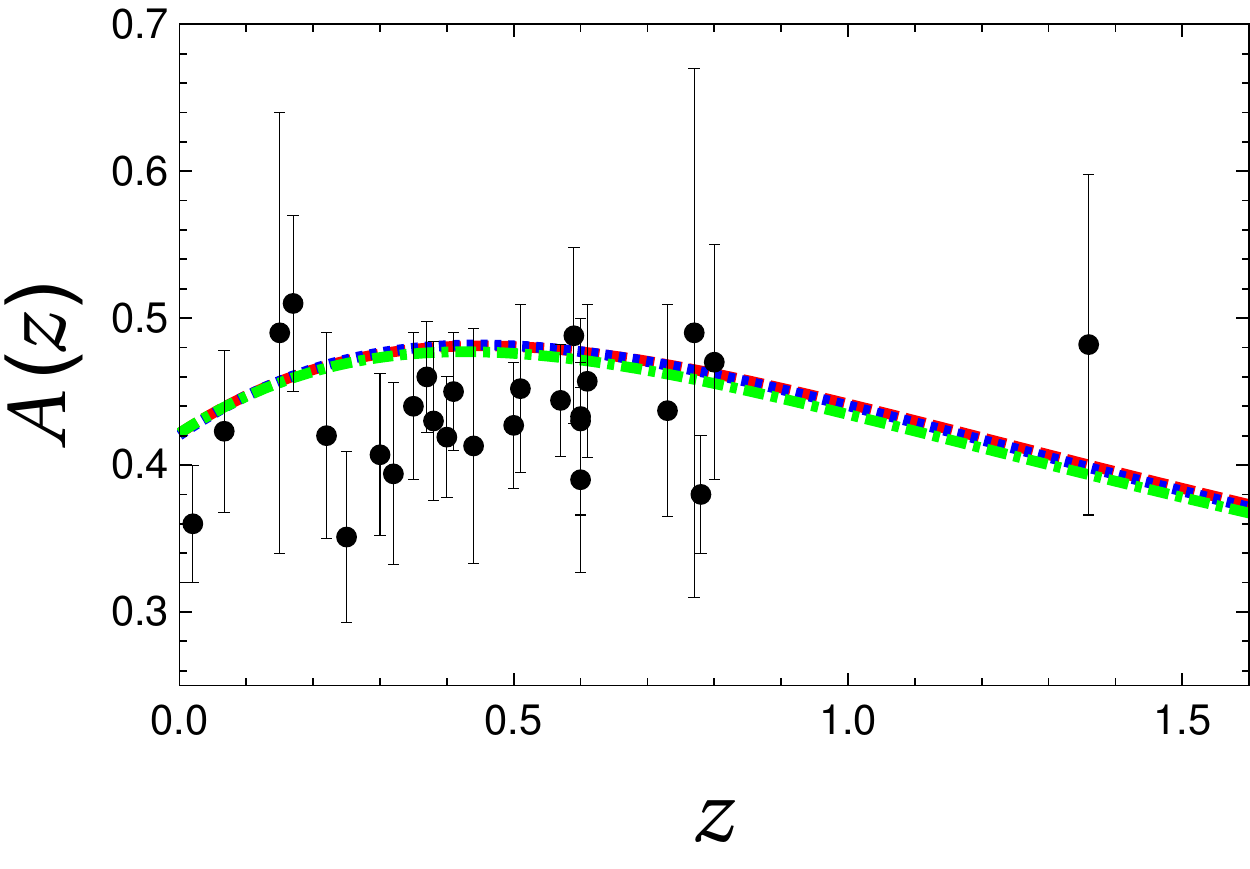}
\caption{\label{fig:4} 
Comparison between observational data (the error bars are shown too) and the prediction of the models I, II and III for $A(z)=\sigma_8(z) f(z)$ versus red-shift for $k = 0.125$ h Mpc$^{-1}$. Shown are: i) model I (dashed red line), ii) model II  (dotted blue line) and iii) model III  (dotted dashed green line).
}
\end{figure}

%
Finally in Fig. \ref{fig:4} we compare the prediction of the models to available data regarding the combination parameter $A(z)=\sigma_8(z) f(z)$, where the rms fluctuation $\sigma_8(z)$ is related to the matter energy density contrast by \cite{Nesseris:2007pa,Albarran:2016mdu}
\begin{equation}
\sigma_8(z)= \frac{\delta_m(z)}{\delta_m(0)} \sigma_8(z=0)
\end{equation}
evaluated at the scale $k_{\sigma_8}=0.125 \: \text{h Mpc}^{-1}$ \cite{Albarran:2016mdu}. We have used for $\sigma_8(z=0)$ the values shown in Table I obtained in \cite{saridakis}. The data points with the error bars as well as the relevant references can be seen in Table II of \cite{Albarran:2016mdu}.
Fig. \ref{fig:4} looks very similar to analogous figures produced in other related works, such as \cite{Albarran:2016mdu,alcaniz,basilakos}.

\begin{table}

\centering
  \caption{Parameters involved in each model}
  \begin{tabular}{cccc}
  \hline
Parameters & Model $\# 1$ & Model $\# 2$ & Model $\# 3$  \\
\hline

$w_0$ 			& -1.0267 &  -1.0517 & -1.0079 \\
$b$ 				& -0.2601 &   0.0113 &  0.1542 \\
$\Omega_{m,0}$ 	    &  0.298  &   0.297  &  0.301  \\
$H_0$	            &  68.95  &   69.02  &  68.59  \\
$\sigma_8$ 			&  0.824  &   0.823  &  0.821  \\

\hline
\end{tabular}
\label{tab:ppnfr}
\end{table}

%
%

%

%
%
%

%
%
%
%
%
%

\section{Conclusions}

To summarize, in this work we have studied the Cosmology of three Oscillating Dark Energy models, both at the level of background evolution and at the level of linear cosmological perturbations. The dark energy equations-of-state were recently studied in the literature, and the free parameters of each model were determined upon comparison against several observational data. First we computed the statefinder parameters $\{r, s\}$ versus red-shift, and after that we computed $f \equiv (a/\delta_m) d \delta_m/da$ as well as the growth index $\gamma \equiv \ln(f) / \ln(\Omega_m)$ as functions of the red-shift for all three dark energy parameterizations, and for three different scales $k$. Our main numerical results are summarized in Fig.~1-4, where the comparison with the $\Lambda$CDM model is shown as well.


\begin{acknowledgments}

We wish to thank the anonymous reviewer for his/her comments and suggestions, and E. Saridakis for correspondence. The work of A.R. was supported by the CONICYT-PCHA/Doctorado Nacional/2015-21151658. G.P. thanks the Funda\c c\~ao para a Ci\^encia e Tecnologia (FCT), Portugal, for the financial support to the Center for Astrophysics and Gravitation-CENTRA, Instituto Superior T\'ecnico, Universidade de Lisboa, through the Grant No. UID/FIS/00099/2013.
\end{acknowledgments}


\end{document}